# Some Aspects of FELWI

K.S. Badikyan

National University of Architecture and Construction, Yerevan, Armenia

badikyan.kar@gmail.com

**Abstract**

A threshold condition for amplification without inversion in a free-electron laser without inversion (FELWI) is determined. This condition is found to be too severe for the effect to be observed in an earlier suggested scheme because a threshold intensity of the field to be amplified appears to be too high. This indicates that alternative schemes have to be found for making the creation of an FELWI realistic.

## 1. Introduction

Usually free-electron lasers (FELs) [1, 2] use the kinetic energy of relativistic electrons moving through a spatially modulated magnetic field (wiggler) to produce coherent radiation. The frequency of radiation is determined by the energy of electrons, the spatial period of magnetic field and the magnetic field strength of the wiggler. This permits tuning a FEL in a wide range unlike atomic or molecular lasers. There are many types of FEL [3-10] and references therein.

However for purposes of achieving a short-wavelength region of generation there are important possible limitations of the FEL gain. The idea of inversionless FEL or FELWI (FEL without inversion) was formulated and discussed in [11–14].
According to the realization [11], of FELWI is strongly related to a deviation of electrons from their original direction of motion owing to interaction with the fields of an undulator and co-propagating light wave. The deviation angle appears to be proportional to energy gained or lost by an electron during its passage through the undulator. Owing to this, a subsequent regrouping of electrons over angles provides regrouping over energies. In principle, a proper installation of magnetic lenses and turning magnets after the first undulator in an FELWI can be used in this case for making faster electrons running over a longer trajectory than the slower ones [12]. This is the *negative-dispersion* condition that is necessary for getting *amplification without inversion* [13].

It is clear that the described mechanism can work only if the interaction-induced deviation of electrons (with a characteristic angle $\alpha$) is larger than the natural angular width beam of the electron

$$\alpha > \alpha_{beam} \tag{1}$$

As the energy gained/lost by electrons in the undulator and the deviation angle are proportional to the field strength amplitude of the light wave to be amplified, condition (1) determines the threshold light intensity, only above which amplification without inversion can become possible. This threshold intensity is estimated below.

## 2. Single particle approximation

In the noncollinear FEL [2-6] the electron slow-motion phase is defined as

$$\varphi = qz + \mathbf{k}\mathbf{r} - \omega t \tag{2}$$

where $q = 2\pi/\lambda_0$ and $\lambda_0$ is the undulator period, $\mathbf{k}$ and $\omega$ are the wave vector and frequency of the wave to be amplified, $|\mathbf{k}| = \omega/c$, $\mathbf{r} = \mathbf{r}(t)$ is the electron position vector and $z = z(t)$ is its projection on the undulator axis. Let the initial electron velocity $\mathbf{v}_0$ be directed along the undulator axis $Oz$. Let the undulator magnetic field $\mathbf{H}$ be directed along the $x$-axis. Let the light wave vector $\mathbf{k}$ be lying on the $(xz)$ plane under an angle $\theta$ to the $z$-axis. Let the electric field strength $\mathbf{\varepsilon}$ of the wave to be amplified be directed along the $y$-axis, as well as its vector potential $\mathbf{A}_{wave}$ wave and the undulator vector potential $\mathbf{A}_{und}$, where

$$A_{wave} = \frac{c\varepsilon_0}{\omega}\cos(\mathbf{k}\mathbf{r} - \omega t), \qquad A_{und} = \frac{H_0}{q}\cos qz, \tag{3}$$

and $\varepsilon_0$ and $H_0$ are the amplitudes of the electric component of the light field and of the undulator magnetic field. The geometry corresponds to that in [1].

The slow motion phase (2) obeys the usual pendulum equation

$$\ddot{\varphi} = -a^2 \sin\varphi, \tag{4}$$

Where

$$a = \frac{ce\sqrt{\varepsilon_0 H_0}}{E_0}, \tag{5}$$

$E_0 = \gamma mc^2$ is the initial electron energy and $\gamma$ is the relativistic factor. If $L$ is the undulator length, the ratio $L/c$ is the time it takes for an electron to pass through the undulator.

The product of this time and the parameter $a$ of equation (5) is known [2, 7-10] as the saturation parameter $\mu$,

$$\mu = \frac{aL}{c} = \frac{eL\sqrt{\varepsilon_0 H_0}}{E_0}. \tag{6}$$

Amplification in an FEL (with $H_0 = const$) is an efficient one as long as $\mu \leq 1$. At $\mu > 1$ the FEL gain $G$ falls. The condition $\mu \sim 1$ determines the saturation field $\varepsilon_{0,sat}$ and intensity $I_{sat}$. For example, at $L = 3m$, $H_0 = 10^4 Oe$, $\gamma = 10^2$, we have $\varepsilon_{0,sat} \sim 1.2 \times 10^2 V/cm$ and $I_{sat} \sim 2 \times 10^5 W/cm^2$. In our further estimates of the FELWI threshold field and intensity we will have to keep in mind that it is hardly reasonable to consider fields stronger than the saturation field $\varepsilon_{0,sat}$.

The pendulum equation (4) has the first integral of motion (kinetic + potential energy of a pendulum = const).

$$\frac{\dot{\varphi}^2}{2} - a^2 \cos[\varphi(t)] = \text{const} \tag{7}$$

Initial conditions of equations (4) and/or (7) are given by

$$\varphi(0) = \varphi_0 \qquad \dot{\varphi}(0) = \delta \equiv \frac{\omega - \omega_{res}}{2\gamma^2} \tag{8}$$

where $\varphi_0$ is an arbitrary initial phase, $\delta$ is the resonance detuning and $\omega_{res}$ is the resonance frequency for noncollinear FEL given by

$$\omega_{res} = \frac{cq}{1-(v_0/c)\cos\theta} \approx \frac{2\gamma^2 cq}{1+\gamma^2\theta^2} \tag{9}$$

with $\theta = angle(\mathbf{k}, Oz)$.

In the case of a not too long undulator and sufficiently small energy width of the electron beam, a characteristic value of the detuning is evaluated as $|\delta| \sim 1/t \sim c/L$.

The rate of change of the electron energy is defined as the work produced by the light field per unit time, and as it is well known [4], this rate is connected directly with the second derivative of the slow-motion phase

$$\frac{dE}{dt} = \frac{E}{2cq}\ddot{\varphi} \approx \frac{E_0}{2cq}\ddot{\varphi}. \tag{10}$$

The last approximate expression is written down in the approximation of a small change of the electron energy, $|E - E_0| \ll E_0$. In this approximation, equation (10) gives the following expression for the total gained or lost energy of a single electron after a passage through the undulator:

$$\Delta E = E\left(\frac{L}{c}\right) - E_0 \approx \frac{E_0}{2cq}\left[\dot{\varphi}\left(\frac{L}{c}\right) - \delta\right]. \tag{11}$$

In the weak-field approximation ($\mu \ll 1$) one can use the iteration method with respect to $a$ of equation (5) for solving equation (7). The zero-order solution is evident and very simple: $\dot{\varphi}^{(0)} \equiv \delta$. In the first order in $a^2$ one obtains

$$\dot{\varphi}^{(1)} = \frac{a^2}{\delta}\left(\cos(\varphi_0 + \delta \cdot t) - \cos(\varphi_0)\right) \sim \frac{a^2 L}{c} = \frac{\mu^2 c}{L}. \tag{12}$$

By substituting this expression into equation (11) we find the first-order change of the electron energy

$$\Delta E^{(1)} = \frac{E_0}{2cq}\dot{\varphi}^{(1)} \sim \frac{E_0}{2cq}\frac{\mu^2 c}{L} = \mu^2 E_0 \frac{\lambda_0}{4\pi L}. \tag{13}$$

Of course, both $\dot{\varphi}^{(1)}$ and $\Delta E^{(1)}$ turn zero being averaged over an arbitrary initial phase $\varphi_0$. But here we are interested in maximal achievable rather than mean values of these quantities, and these maximal values are given just by estimates of equations (12) and (13).

In accordance with the results of [13] and [14] a transverse velocity $v_x$ and energy $\Delta E$ acquired by an electron after a passage through the undulator are directly proportional to each other

$$v_x = c\theta \frac{\Delta E}{E_0}, \qquad (14)$$

which gives in the first order the following estimate of the electron deviation angle $\alpha$:

$$\alpha \approx \frac{v_x^{(1)}}{v_0} \approx \frac{v_x^{(1)}}{c} = \theta \frac{\Delta E^{(1)}}{E_0} \sim \theta \mu^2 \frac{\lambda_0}{4\pi L} \sim \mu^2 \frac{d\lambda_0}{4\pi L^2}, \qquad (15)$$

where $d$ is the electron beam diameter and we took $\theta = d/L$.

As said above, in the framework of a linear theory we can consider only such fields at which $\mu \leq 1$. Moreover, consideration of the case $\mu \gg 1$ has no sense at all, because the corresponding fields are too strong and because saturation makes the gain too small. For these reasons let us take for
estimates the maximal value of the saturation parameter μ compatible with the weak-field approximation, $\mu \sim 1$. Let us take also $\lambda_0 = 3 cm$, $d = 0.3$ cm and $L = 3\times 10^2 cm$. Then, we get from equation (15) the following estimate of the electron deviation angle:

$$\alpha \sim 10^{-6}. \qquad (16)$$

At weaker fields and smaller values of the saturation parameter μ the deviation angle $\alpha$ is even smaller than that given by equation (16). But even at μ = 1 the angle $\alpha$ is very small.

## 3. Conclusion

To make the estimate (16) compatible with the condition of equation (1) one has to provide a natural electron beam angular divergence smaller than $10^{-6}$. Unfortunately, such weakly diverging electron beams hardly exist. Hence, the creation of an FELWI requires invention of alternative schemes in which threshold restrictions would be much weaker than in the considered one.